\documentclass[11pt]{article}
\pdfoutput=1
\usepackage[pdftex]{graphicx}
\usepackage{epsfig}
\usepackage{epstopdf}
\usepackage{cite}
\newcommand{\AmS}{{\protect\the\textfont
  A\kern-.1667em\lower.5ex\hbox{M}\kern-.125emS}}
\newcommand{\ba}{\begin{array}}
\newcommand{\ea}{\end{array}}
\newcommand{\lsim}{\raisebox{-0.13cm}{~\shortstack{$<$ \\[-0.07cm] $\sim$}}~} 
\newcommand{\gsim}{\raisebox{-0.13cm}{~\shortstack{$>$ \\[-0.07cm] $\sim$}}~} 
\def\beq{\begin{equation}}   
\def\eeq{\end{equation}}

\def\nnum{\nonumber}
\def\bea{\begin{eqnarray}}
\def\eea{\end{eqnarray}}
\def\noi{\noindent}
\def\cl{{\rm c}}
\def\sg{{\rm s}}
\def\bg{{\rm b}}

\def\beq{\begin{equation}}   
\def\eeq{\end{equation}}

\def\nnum{\nonumber}
\def\bea{\begin{eqnarray}}
\def\eea{\end{eqnarray}}
\def\ct{\cite}

\begin{document}
\begin{titlepage}
\begin{flushright}
Report: IFIC/17-27, FTUV-17-0609\\
\end{flushright}

\begin{center}
\vspace{2.7cm}
{\Large{\bf 
The ridge effect and three-particle correlations}}
\end{center}

\vspace{1cm}

\begin{center}

{\bf Miguel-Angel 
Sanchis-Lozano$^{\rm a,}$\footnote{Email 
address: Miguel.Angel.Sanchis@ific.uv.es}, 
Edward Sarkisyan-Grinbaum$^{\rm b,c,}$\footnote{Email address: 
Edward.Sarkisyan-Grinbaum@cern.ch}
\vspace{1.5cm}\\
\it 

\it $^{\rm a}$ Instituto de F\'{\i}sica
Corpuscular (IFIC) and Departamento de F\'{\i}sica Te\'orica \\
\it Centro Mixto Universitat de Val\`encia-CSIC, 
Dr. Moliner 50, E-46100 Burjassot, Valencia, Spain
\\ 
$^{\rm b}$ 
Experimental Physics Department, CERN, 1211 Geneva 23, Switzerland\\
$^{\rm c}$ Department of Physics, The University of Texas at Arlington,
Arlington, 
TX 
76019, USA}

\end{center}

\vspace{0.5cm}

\begin{abstract}
 Pseudorapidity and azimuthal three-particle correlations are studied
based on a correlated-cluster model of multiparticle production. The model
provides a common framework for correlations in proton-proton and
 heavy-ion collisions allowing
 easy
 comparison with the measurements. It
is shown that azimuthal cluster correlations are definitely required in
order to understand three-particle correlations in the near-side ridge
effect.
This is similar to the explanation of the ridge phenomenon found
in our previous analysis of two-particle correlations and
generalizes the model to
 higher-order correlations.
 \end{abstract}

\begin{center}


\end{center}

\end{titlepage}


\section{Introduction}

 Correlation measurements 
have proven to be extremely useful to understand multiparticle production 
providing stringent tests for theoretical models of soft hadronic dynamics 
\ct{book,DeWolf:1995pc,vhm,Dremin:2000ep}.
  Being sensitive to the last stage of particle interactions (freeze-out) 
in hadronic collisions, correlations become of special interest in 
high-energy heavy-ion collisions for the study of new matter formation, 
 such as quark-gluon plasma and its properties \ct{CollEff-rev}, 
 as well as in 
 the 
search for hidden sectors
beyond the Standard Model \cite{Sanchis-Lozano:2015eca,SanchisLozano:2008te}.

Moreover, somewhat unexpected collective 
effects have been observed both 
in heavy-ion 
and proton-proton collisions at RHIC and LHC experiments  
\cite{CollEff-rev}. Striking 
ridge-like and dip structures show up in the two-particle 
correlation spectrum. In particular, the near-side ridge phenomenon 
corresponds to azimuthal collimated particle production extending over a 
large pseudorapidity interval, whose physical origin is still unclear 
especially for proton-proton collisions where no collective flow is 
 expected.

In a previous paper 
\cite{Sanchis-Lozano:2016qda} 
we studied 
two-particle correlations for the analysis of the near-side ridge in 
hadron-hadron collisions. The effect 
was shown to have a natural explanation 
provided that clusters are produced 
in a correlated way in the collision transverse plane.
 On the other hand, if particles are emitted correlated to each other, 
this effect should also hold for higher-rank particle correlations. Therefore, 
the analysis of three-particle correlations 
 attracts  as well 
 high interest 
   providing further information on
 hadroproduction mechanisms 
 \ct{book,DeWolf:1995pc,vhm,Dremin:2000ep}.

In this work we extend our study \cite{Sanchis-Lozano:2016qda} 
to three-particle (pseudo)rapidity and 
 azimuthal correlations in the framework of a correlated-cluster model (CCM), 
   providing compact
 final expressions with 
predictions under some simple physical assumptions, which can be directly 
  tested by experiments.

 \section{Definitions and notations}
As usual, two-particle correlations can be studied by means of
\begin{equation}\label{eq:C2}
C_2(1,2)=\rho_2(1,2)-\rho(1)\rho(2)\, ,
\label{eq:corfunction}
\end{equation}
where indices $1$ and $2$ stand for the set of kinematic variables relative to 
particles $1$ and $2$, respectively. In terms of the rapidity ($y$) and 
 the azimuthal angle ($\phi$), the one-particle density  $\rho$ and the two-particle density 
$\rho_2$ are defined through
\beq\label{eq:def2}
\rho(y,\phi)=\frac{1}{\sigma_{\rm 
in}}\frac{d^2\sigma}{dyd\phi} \ ;\  \ \rho(y_1,y_2,\phi_1,\phi_2) = 
\frac{1}{\sigma_{\rm in}}\frac{d^4\sigma}{dy_1d\phi_1dy_2d\phi_2} \, , 
\eeq
where $\sigma_{\rm in}$ denotes the inelastic cross section and
the dependence on the transverse momentum ($p_T$) has been integrated out. Still
a potential dependence on the $p_T$ integration range should remain, depending on the selected
kinematic cuts.

The three-particle rapidity correlation function is defined as:
\beq\label{eq:3-Cor}
C_3(1,2,3)= \rho_3(1,2,3)+2\rho(1)\rho(2)\rho(3)-\rho_2(1,2)\rho(3)-
\rho_2(2,3)\rho(1)-\rho_2(1,3)\rho(2)\, ,
\eeq
where the three-particle density is defined as
\begin{equation}\label{eq:def3}
\rho_3(y_1,y_2,y_3,\phi_1,\phi_2,\phi_3)\ =\
\frac{1}{\sigma_{\rm in}}\ 
\frac{d^6\sigma}{dy_1dy_2dy_3d\phi_1d\phi_2d\phi_3}\, . 
\end{equation}

As is well known, three-particle correlations can provide a more 
sophisticated test than two-particle correlations for the study of 
partonic dynamics (e.g. jets) in proton-proton collisions 
\cite{Ramos:2011tw},
 or the emergence of a new state of matter 
in heavy-ion collisions \cite{He:2017zpg}.

\subsection{Correlations as function of (pseudo)rapidity and azimuthal 
differences}

In order to match our theoretical approach to experimental results in terms of 
(pseudo)rapidity and azimuthal differences ($\Delta y_{ij}=y_i-y_j$ and 
$\Delta \phi_{ij}=\phi_i-\phi_j$, $i,j=1,2,3,\ i\neq j$), 
use will be made of Dirac's  $\delta$-functions as 
 in  our earlier studies  
\cite{Sanchis-Lozano:2016qda}. Then, the
two-particle distribution of uncorrelated pairs reads
\begin{equation}
b_2(\Delta y_{ij}, \Delta \phi_{ij})=\int dy_idy_jd\phi_id\phi_j\ 
\rho(y_i,\phi_i)\ \rho(y_j,\phi_j)\ \delta(\Delta y_{ij} -y_i+y_j)\
\delta(\Delta \phi_{ij} -\phi_i+\phi_j)  \, ,
\end{equation}
and the distribution of correlated pairs can be identified with
 \begin{equation}
   s_2(\Delta y_{ij}, \Delta \phi_{ij})=\int dy_idy_j d\phi_id\phi_j\  
\rho_2(y_i,\phi_i,y_j,\phi_j)\ \delta(\Delta y_{ij} -y_i+y_j)\
\delta(\Delta \phi_{ij} -\phi_i+\phi_j) \, .
\end{equation}

Three-particle correlations are again expressed as a function of the 
rapidity and azimuthal differences,\footnote{Notice that only two rapidity 
differences are independent: $\Delta y_{12}=y_1-y_3$ and $\Delta 
y_{13}=y_1-y_3$ are chosen as independent, so that $\Delta 
y_{23}=y_2-y_3=\Delta y_{13}-\Delta y_{12}$. Similarly for the azimuthal 
variable:  $\Delta \phi_{12}=\phi_1-\phi_2$ and $\Delta 
\phi_{13}=\phi_1-\phi_3$ are independent, so that $\Delta 
\phi_{23}=\phi_2-\phi_3=\Delta \phi_{13}-\Delta \phi_{12}$.}

\begin{equation}\label{eq:s3}
s_3(\vec{\Delta y}, \vec{\Delta \phi})\ =\ 
\int d\vec{y}\ d\vec{\phi}\ \vec{\delta}(\Delta y)\ \vec{\delta}(\Delta 
\phi)\ \rho_3(\vec{y},\vec{\phi})\, ,
\end{equation}
where the shortened notation has been introduced:

\[  
\vec{\Delta y}, \vec{\Delta \phi}\,\;  {\rm  for}\,\; \Delta y_{ij}, 
\Delta 
\phi_{ij}\ ,\,
\vec{y}=(y_1, y_2, y_3)\ ,\ 
\vec{\phi} = (\phi_1, \phi_2, \phi_3)\ ,\ 
d\vec{y}\ d\vec{\phi}\ = 
dy_1dy_2dy_2\ d\phi_1d\phi_2d\phi_3\, ,
\] 
  and for the Dirac's $\delta$-functions: 
\begin{equation}\label{eq:dirac3}
 \vec{\delta}(\Delta y) = \delta(\Delta y_{12}-y_1+y_2)\ 
\delta(\Delta y_{13}-y_1+y_3), 
\end{equation}
\[ \vec{\delta}(\Delta \phi) = \delta(\Delta \phi_{12}-\phi_1+\phi_2)\ 
\delta(\Delta \phi_{13}-\phi_1+\phi_3)\, . 
\]

Non-correlated three-particle distributions correspond to 
\begin{equation}\label{eq:b3}
b_3(\vec{\Delta y}, \vec{\Delta \phi})\ =\ 
\int  d\vec{y}\ d\vec{\phi}\ \vec{\delta}(\Delta y)\ \vec{\delta}(\Delta 
\phi)\ \rho(y_1,\phi_1)\ \rho(y_2,\phi_2)\ \rho(y_3,\phi_3)\, . 
\eeq

According to Eq.(\ref{eq:3-Cor}), a three-particle normalized correlation 
function 
depending on the rapidity and
azimuthal differences can be defined as
\begin{equation}\label{eq:3-cor}
c_3(\vec{\Delta y},\vec{\Delta \phi})= 
\frac{s_3+2b_3-s_{123}-s_{231}-s_{132}}{b_3}\, ,
\end{equation}
where the explicit dependence on the rapidity and azimuthal diferences has 
been omitted in the 
terms of the 
     r.h.s., and

\begin{equation}\label{eq:eq:s123}
s_{123}(\vec{\Delta y}, \vec{\Delta \phi})\ =\ 
\int d\vec{y}\ d\vec{\phi}\ \vec{\delta}(\Delta y)\ \vec{\delta}(\Delta 
\phi)\ \rho(y_1,\phi_1)\ 
\rho_2(y_2,\phi_2,y_3,\phi_3)\, , 
\eeq
    while 
 $s_{231}$ and $s_{132}$ terms are obtained straigthforwardly by permutation.

On the other hand, sometimes a simplified version of the three-particle 
correlation function, also of common use in experimental 
analyses of data, is given by  \ct{CollEff-rev}
\begin{equation}\label{eq:3-cor-short}
c_3(\vec{\Delta y},\vec{\Delta \phi})= \frac{s_3}{b_3}\, .
\end{equation}

In the following we make use of the expression (\ref {eq:3-cor-short}) 
although the main conclusions would remain the same had we 
employed Eq.(\ref {eq:3-cor}) instead.\footnote{Additionaly, 
  correlations among so-called event planes \cite{Bhalerao:2013ina} 
(corresponding to different harmonics) have recently emerged as a powerful 
tool for the analysis of heavy-ion collisions 
\cite{Adamczyk:2017hdl,Adamczyk:2017byf}. In this paper, which can be 
applied to proton-proton collisions as well, we do not consider this 
analysis.}

 \section{Two- and three-particle correlations in the CCM}

 It is 
 generally 
 accepted that particle production in soft hadronic interactions 
occurs
via an intermediate step of decaying strings/clusters/fireballs yielding 
final-state particles \cite{book,Dremin:2000ep}. 
It should be noted that the ``cluster'' 
 concept has to be understood
in a broad sense, i.e. a group of particles with some correlated 
properties, probably comming from a common ancestor.

We keep the same notation as in our paper  \cite{Sanchis-Lozano:2016qda}
for two-particle correlations. Hence, 
the single particle density can be expressed as the convolution of the 
cluster density $\rho^{(\cl)}(y_\cl,\phi_\cl)$
and the particle density from a single cluster $\rho^{(1)}(y,\phi;y_\cl,\phi_\cl)$, i.e.
\beq\label{eq:rho}
\rho(y,\phi)=\int dy_\cl d\phi_\cl\ \rho^{(\cl)}(y_\cl,\phi_\cl)\ 
\rho^{(1)}(y,\phi;y_\cl,\phi_\cl)=\langle N_\cl\rangle\ \bar{\rho}^{(1)}\ E_1(y,\phi) \, , \int dy\ d\phi\ E_1(y,\phi)=1 \, .
\eeq
where $\langle N_{\cl} \rangle$ stands for the average cluster number 
per collision and
$\bar{\rho}^{(1)}$ denotes the average particle density for single cluster decays. On the other hand,  
the function $E_1(y,\phi)$ encodes the 
expected dependence on the rapidity and azimuthal variables of the emitted 
particles.

For uncorrelated particle pairs and triplets  we introduce the product 
of the two 
and three single-particle distributions 
representing the mixed-event background, 
\beq\label{eq:rho2b}
\rho_{\rm mixed}(y_1,\phi_1,y_2,\phi_2)\ =\ 
\rho(y_1,\phi_1)\rho(y_2,\phi_2)\ =\ \langle N_\cl \rangle^2\ 
\bar{\rho}^{(1)2}E_1(y_1,\phi_1)E_1(y_2,\phi_2) \, ,
\eeq
\beq\label{eq:rho3b}
\rho_{\rm mixed}(\vec{y},\vec{\phi})\ =\ 
\rho(y_1,\phi_1)\rho(y_2,\phi_2)\rho(y_3,\phi_3)\ =\ \langle N_\cl 
\rangle^3\ \bar{\rho}^{(1)3}E_1(y_1,\phi_1)E_1(y_2,\phi_2)E_1(y_3,\phi_3) 
\, ,
\eeq
which suggests to define
\beq\label{eq:Eb2}
E_\bg (y_1,\phi_1,y_2,\phi_2)\ =\ E_1(y_1,\phi_1)E_1(y_2,\phi_2)\, ,
\eeq
\beq\label{eq:Eb3}
E_\bg (\vec{y},\vec{\phi})\ =\ E_1(y_1,\phi_1)E_1(y_2,\phi_2)E_1(y_3,\phi_3) \, .
\eeq

Next, the two-particle density can be written as
\beq\label{eq:rho2}
\rho_2(y_1,\phi_1,y_2,\phi_2)\ =\  
\int dy_\cl \phi_\cl\ \rho^{(\cl)}(y_\cl,\phi_\cl)\ 
\rho_2^{(1)}(y_1,\phi_1,y_2,\phi_2;y_\cl,\phi_\cl)\ 
\eeq
\[
+\ \int dy_{\cl 1}dy_{\cl 2}d\phi_{\cl 1}d\phi_{\cl 2}\ 
\rho_2^{(\cl)}(y_{\cl 1},\phi_{\cl 1},y_{\cl 2},\phi_{\cl 2})\ 
\rho^{(1)}(y_1,\phi_1;y_{\cl 1},\phi_{\cl 1})\  
\rho^{(1)}(y_2,\phi_2;y_{\cl 2},\phi_{\cl 2}) \, .
\]
The first term on the r.h.s. corresponds to the emission 
of secondaries from a single cluster while the second term corresponds to
the emission of the two particles from two distinct clusters, whose
density is noted as $\rho_2^{(\cl)}(y_{\cl 1},\phi_{\cl 1},y_{\cl 2},\phi_{\cl 2})$. Therefore, we conclude for the two-particle density:
\beq\label{eq:corre2}
\rho_2(y_1,\phi_1,y_2,\phi_2)\ =\  
\langle N_\cl \rangle\ \bar{\rho}^{(1)2} 
E_\sg^{(1)}(y_1,\phi_1,y_2,\phi_2)\ +\ \langle N_\cl(N_\cl-1) \rangle\ 
\bar{\rho}^{(1)2}E_\sg^{(2)}(y_1,\phi_1,y_2,\phi_2) \, , 
\end{equation}
where $E_\sg^{(1)}(y_1,\phi_1,y_2,\phi_2)$ and 
$E_\sg^{2}(y_1,\phi_1,y_2,\phi_2)$ 
stand for correlations stemming from 
the corresponding two integrals of Eq.(\ref{eq:rho2}).

In its turn, the three-particle density can be written as
\beq\label{rho3}
\rho_3(\vec{y},\vec{\phi})= 
\int dy_\cl d\phi_\cl\ \rho^{(\cl)}(y_\cl,\phi_\cl)\ 
\rho_3^{(1)}(\vec{y},\vec{\phi};y_\cl,\phi_\cl)\ 
\eeq
\[ 
+\ \int dy_{\cl 1}dy_{\cl 2}d\phi_{\cl 1}d\phi_{\cl 2}\ 
\rho_2^{(\cl)}(y_{\cl 1},\phi_{\cl 1})\ \rho^{(\cl)}(y_{\cl 2},\phi_{\cl 
2})\ 
\rho^{(2)}(y_1,\phi_1;y_{\cl 1},\phi_{\cl 1})\ 
\rho^{(1)}(y_2,\phi_2;y_{\cl 2},\phi_{\cl 2})\ +\ {\rm permutations}\ 
\]
\[
+\ \int dy_{\cl 1}dy_{\cl 2}dy_{\cl 3}d\phi_{\cl 1}d\phi_{\cl 
2}d\phi_{\cl 3}\ 
\rho_3^{(\cl)}(\vec{y}_\cl,\vec{\phi}_\cl)\ 
\rho^{(1)}(y_1,\phi_1;y_{\cl 1},\phi_{\cl 1})\ 
\rho^{(1)}(y_2,\phi_2;y_{\cl 2},\phi_{\cl 2})\ 
\rho^{(1)}(y_3,\phi_3;y_{\cl 3},\phi_{\cl 3})\, ,
\]
where we have introduced $\vec{y}_\cl \equiv (y_{\cl 1},y_{\cl 2},y_{\cl 
3})$ 
and 
$\vec{\phi}_\cl \equiv (\phi_{\cl 1},\phi_{\cl 2},\phi_{\cl 3})$. We will 
write
\beq\label{eq:corre3}
\rho_3(\vec{y},\vec{\phi})\ =\  
\langle N_\cl \rangle\ \bar{\rho}^{(1)3} 
E_\sg^{(1)}(\vec{y},\vec{\phi})\ +\ \langle N_\cl(N_\cl-1) \rangle\ 
\bar{\rho}^{(1)3}E_\sg^{(2)}(\vec{y},\vec{\phi})\ +\ 
\langle N_\cl(N_\cl-1) (N_\cl-2)\rangle\ 
\bar{\rho}^{(1)3}E_\sg^{(3)}(\vec{y},\vec{\phi})
\end{equation}
where the functions
$E_\sg^{(k)}(\vec{y},\vec{\phi})$, $k=1,2,3$,   
encode the rapidity and angular dependence
for three-particle correlations in single-cluster production ($k=1$), 
double-cluster production ($k=2$), and
triple-cluster production ($k=3$).

\subsection{Factorization hypothesis}

As in \cite{Sanchis-Lozano:2016qda}, we apply factorization of the
rapidity (longitudinal) and
azimuthal (transverse) directions to the $E$-functions.
Factorization of production cross sections and decay distributions 
into transverse and longitudinal momentum parts is a hypothesis
widely used in many high-energy physics processes. Although 
not yet rigorioulsy proven from first principles, it 
works very well when contrasted 
with experimental data, 
especially for high transverse momentum where such a hypothesis
can also be theoretically justified to some extent.

Since the ridge phenomenon shows up for particles with transverse 
momentum typically of order $\gsim 1$  GeV,
 we will factorize the 
 rapidity and 
 azimuthal dependences of the above $E$-functions 
following \cite{Sanchis-Lozano:2016qda}, as
\bea\label{eq:factor}
E_\bg (\vec{y},\vec{\phi}) & = & E_\bg ^L(\vec{y}) \cdot 
E_\bg ^T(\vec{\phi}) \nnum \, , \\
E_\sg(\vec{y},\vec{\phi}) & = & E_\sg^L(\vec{y}) \cdot 
E_\sg^T(\vec{\phi}) \, ,
\eea
where the superscripts $L$ and $T$ denote the longitudinal 
and transverse parts, respectively.
 
According to different (hydrodynamic, cascade) models, fluctuating initial conditions should lead 
to decorrelations of the orientation of initial event-planes in 
heavy ion collisions. In particular, the authors of \cite{Bozek:2015bna,Bozek:2015bha} argue that event-to-event early state fluctuations (termed $\lq\lq$torque effect'' in \cite{Moreira:2011ht}) should lead to a
(pseudo)rapidity-azimuthal factorization breaking for well separated (pseudo)rapidity 
particles in heavy ion collisions. Notice, however, that keeping the (pseudo)rapidity difference 
$\Delta y_{ij}$ small, Eqs.(\ref{eq:factor}) should remain reliable. 
Moreover, any observed deviation from our later predictions on rapidity and azimuthal correlations
for larger (pseudo)rapidity separations might be
interpreted as a hint of the existence of such kind of {\em torque effect}.

On the other hand, as usual in cluster models we shall adopt Gaussian distributions  
in rapidity and azimuthal spaces for both cluster density 
and particle density from clusters, as developed below. Thus, we shall 
write the single,
two-cluster 
and three-cluster 
densities as
\[
\rho^{(\cl)}(y_\cl,\phi_\cl) \sim  
\exp{\left[-\frac{y_\cl^2}{2\delta_{\cl y}^2}\right]}\, ,\ 
\rho_2^{(\cl)}(y_{\cl 1},\phi_{\cl 1},y_{\cl 2},\phi_{\cl 2}) \sim  
\exp{\left[-\frac{(y_{\cl 1}+y_{\cl 2})^2}{2\delta_{\cl y}^2}\right]}  
\times
\exp{\left[-\frac{(\phi_{\cl 1}-\phi_{\cl 2})^2}{2\delta_{\cl 
\phi}^2}\right]}\, ,
\]
\beq\label{eq:clustercorr}
\rho_3^{(\cl)}(\vec{y_{\cl}},\vec{\phi_{\cl }}) \sim 
\exp{\left[-\frac{(y_{\cl 1}+y_{\cl 2}+y_{\cl 3})^2}{2\delta_{\cl 
y}^2}\right]} 
\times
\exp{\left[-\frac{(\phi_{\cl 1}-\phi_{\cl 2})^2+(\phi_{\cl 1}-\phi_{\cl 
3})^2+(\phi_{\cl 2}-\phi_{\cl 3})^2
}{2\delta_{\cl \phi}^2}\right]}\, ,
\eeq
where $\delta_{\cl y}$ and $\delta_{\cl \phi}$ stand for the rapidity and 
azimuthal cluster correlation lengths, respectively.
  Let us remark that Eqs. (\ref{eq:clustercorr}) 
can be regarded as parameterizations especially 
suitable to determine the near-ridge effect using the CCM. 
 %
 The rapidity Gaussians with arguments
 $y_{\cl 1}+y_{\cl 2}$ and $y_{\cl 1}+y_{\cl 2}+y_{\cl 3}$
can be seen as a consequence of (partial) longitudinal 
momentum conservation for two-cluster and three-cluster poduction.   
The azimuthal conditions are implemented in the
Gaussians following \cite{Sanchis-Lozano:2016qda},  
in order to include collinear emission of particles in the 
near-side ridge effect.

On account of the plateau structure
of multiplicity distribution in pseudorapidity phase space, one
may assume that the dependence of $\rho^{(\cl)}(y_\cl,\phi_\cl)$ on
$y_\cl$ is rather weak, i.e. $\delta_{\cl y}^2 \gg 1$. On the other hand, 
the particle density from single cluster decay, the rapidity and azimuthal 
dependence can be approximately expressed in terms of 
Gaussians, i.e.
\beq\label{eq:part-rap-azim}
\rho^{(1)}(y,\phi;y_\cl,\phi_\cl)\ \sim\ 
\exp{\left[-\frac{(y-y_\cl)^2}{2\delta_y^2}\right]} 
\times
\exp{\left[-\frac{(\phi-\phi_\cl)^2}{2\delta_{\phi}^2}\right]} \, .
\eeq 

 The parameter $\delta_y$ ($\lsim 1$ rapidity units \cite{Alver:2007wy}) is 
usually referred to as the cluster decay (pseudo)rapidity
 ``width''. Regarding the transverse plane, $\delta_{\phi}$ can be seen
as another cluster decay width.  For small azimuthal angles with respect 
to the  
cluster direction, $\delta_{\phi}\ \sim\ \frac{1}{v_T\gamma_T}$, where
$v_T$ and $\gamma_T$ denote the cluster velocity in the transverse plane
and its associated Lorentz factor,  
as shown in \cite{Sanchis-Lozano:2016qda}.

\begin{figure}[t]
\begin{center}
\includegraphics[scale=0.33]{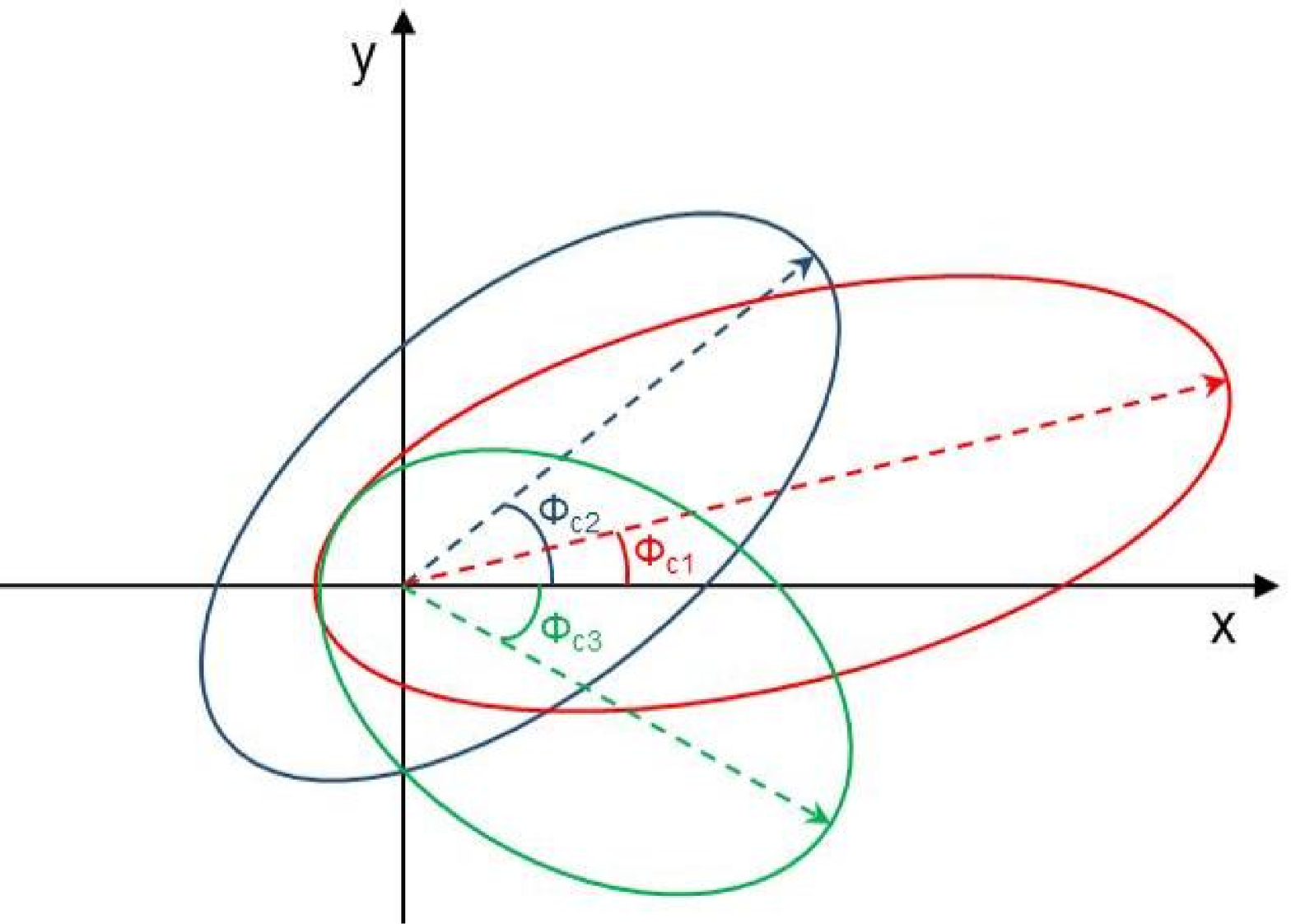}
\caption{Illustrative picture of three clusters produced in a primary 
hadron collision at the origin of the transverse plane 
with azimuthal angles $\phi_{\cl 1}$, $\phi_{\cl 2}$ and $\phi_{\cl 3}$,  
decaying into final-state particles.}
\label{fig:drawing}
\end{center}
\end{figure}

 In Fig. {\ref{fig:drawing}} 
we illustrate the particle emission from three clusters
produced at the same primary hadron collision leading to different
elliptic shapes due to different Lorentz boosts. Clusters are assumed
to be correlated both in rapidity and azimuth according to 
Eqs. (\ref{eq:clustercorr}) (for more details see Appendices).

\section{Interpretation of the near-side ridge effect according to CCM}

 Rewriting the three-particle correlation function by adding the 
terms of contributions from one, two and three 
cluster productions, one gets:
\begin{equation}\label{eq:Cfinal}
c_3(\vec{\Delta y}, \vec{\Delta \phi})=\frac{s_3^{(1)}(\vec{\Delta y},\vec{\Delta \phi})+
s_3^{(2)}(\vec{\Delta y},\vec{\Delta \phi})+
s_3^{(3)}(\vec{\Delta y},\vec{\Delta \phi})}{b_3(\vec{\Delta 
y},\vec{\Delta \phi})}\, ,
\end{equation}
\[
=\frac{1}{\langle N_\cl \rangle^2}\ h^{\rm (1)}(\vec{\Delta y},\vec{\Delta \phi})\  
+\ \frac{\langle N_\cl(N_\cl-1) \rangle}{\langle N_\cl \rangle^3}\ h^{\rm (2)}(\vec{\Delta y},\vec{\Delta \phi})\ +\ 
\frac{\langle N_\cl(N_\cl-1)(N_\cl-2) \rangle}{\langle N_\cl \rangle^3}\ 
h^{\rm (3)}(\vec{\Delta y},\vec{\Delta \phi})\, ,
\]
where detailed expressions for the $h$-functions 
 are given 
 in Appendix \ref{sec:final3}.
 For Poisson
 distribution of clusters,  
$\langle N_\cl(N_\cl-1) 
\rangle$ becomes $\langle N_\cl \rangle^2$ and
 $\langle N_\cl(N_\cl-1)(N_\cl-2) \rangle$ becomes $\langle N_\cl 
\rangle^3$. 
 
Therefore, 
 the 
 above
 expression leads to 
\begin{equation}\label{eq:CPois}
c_3(\vec{\Delta y}, \vec{\Delta \phi})
=\frac{1}{\langle N_\cl \rangle^2}\ h^{\rm (1)}(\vec{\Delta y},\vec{\Delta \phi})\  
+\ \frac{1}{\langle N_\cl \rangle}\ h^{\rm (2)}(\vec{\Delta y},\vec{\Delta \phi})\ 
+\ h^{\rm (3)}(\vec{\Delta y},\vec{\Delta \phi})\, .
\eeq
 The last expression
 shows that 
the $h^{\rm (3)}$ contribution
 dominates for large $\langle N_\cl \rangle$ (hence for high-multiplicity
events).
 This is an important 
 feature
 concerning the correlated-cluster production
as discussed later. 


In the limit $\delta_{\cl y}^2\gg\delta_{y}^2$, $\delta_{\cl 
 \phi}^2\gg \delta_{\phi}^2$ and 
 keeping the
 ($\Delta y_{12}$, $\Delta y_{13}$) and
($\Delta \phi_{12}$, $\Delta \phi_{13}$) 
 components,
 Eqs. (\ref{eq:h1bis})--(\ref{eq:h3bis}) of Appendix \ref{sec:final3} 
 read:\\

\noi
{\it - for one cluster:}
\[
h^{\rm (1)}(\Delta y_{12},\Delta y_{13},\Delta \phi_{12},\Delta \phi_{13})\ 
\sim\ 
\exp{\left[-\frac{(\Delta y_{12})^2+(\Delta y_{13})^2-\Delta y_{12}\Delta 
y_{13}}
{3\delta_y^2} \right]} 
\]
\beq\label{eq:h1}
 \times
\exp{\left[-\frac{(\Delta \phi_{12})^2+(\Delta \phi_{13})^2-2\Delta 
\phi_{12}\Delta \phi_{13}}
{3\delta_{\phi}^2}\right]}\, , 
\eeq

\noi
{\it - for two clusters:}
\[
h^{\rm (2)}(\Delta y_{12},\Delta y_{13},\Delta \phi_{12},\Delta \phi_{13})\ 
\]
\[
\sim\ 
\left( 
\exp{\left[-\frac{(\Delta y_{12})^2}{4\delta_y^2}\right]}+
\exp{\left[-\frac{(\Delta y_{13})^2}{4\delta_y^2}\right]}+
\exp{\left[-\frac{(\Delta y_{12})^2+(\Delta y_{13})^2-2\Delta y_{12}\Delta 
y_{13}}{4\delta_y^2}\right]}
\right)
\]
\[
\times\ 
\left(
\exp{\left[-\frac{(\Delta \phi_{12})^2}{4\delta_{\phi}^2}\right]}+
\exp{\left[-\frac{(\Delta \phi_{13})^2}{4\delta_{\phi}^2}\right]}+ 
\exp{\left[-\frac{(\Delta \phi_{12})^2+(\Delta \phi_{13})^2-2\Delta 
\phi_{12}\Delta \phi_{13}}{4\delta_{\phi}^2}\right]}
\right)\ 
\]
\beq\label{eq:h2} 
\times
\exp{\left[-\frac{(\Delta y_{12})^2+(\Delta y_{13})^2-\Delta 
y_{12}\Delta y_{13}}
{3\delta_{\cl y}^2}\right]}\
\exp{\left[-\frac{(\Delta \phi_{12})^2+(\Delta \phi_{13})^2-\Delta 
\phi_{12}\Delta \phi_{13}}
{2\delta_{\cl \phi}^2}\right]}
\, ,
\eeq

\noi
{\it - for three clusters:}
\beq\label{eq:h3}
h^{\rm (3)}(\Delta y_{12},\Delta y_{13},\Delta \phi_{12},\Delta 
\phi_{13})\,  
\sim\, \exp{\left[\frac{(\Delta y_{12})^2+(\Delta y_{13})^2+-(\Delta 
y_{12})(\Delta y_{13})}{3\delta_{\cl y}^2}\right]}\ 
\eeq
\[
\times\
\left(
\exp{\left[-\frac{(\Delta \phi_{12})^2+(\Delta \phi_{13})^2-\Delta 
\phi_{12}\Delta \phi_{13}}
{\delta_{\cl \phi}^2}\right]} \right.
\]
\[
\left.
+\
\exp{\left[-\frac{(\Delta \phi_{12})^2}{2\delta_{\cl \phi}^2}\right]}+
\exp{\left[-\frac{(\Delta \phi_{13})^2}{2\delta_{\cl \phi}^2}\right]}+ 
\exp{\left[-\frac{(\Delta \phi_{12})^2+(\Delta \phi_{13})^2-2\Delta 
\phi_{12}\Delta \phi_{13}}{2\delta_{\cl \phi}^2}\right]} \right)\, .
\]
\\

  Figures \ref{fig:contour}--\ref{fig:projrap}, 
 based on Eqs. (\ref{eq:CPois})--(\ref{eq:h3}), illustrate the 
above-described interpretation.
  
 In 
 the left panel of  Fig.\ref{fig:contour}
 we show the contour-plot of
$c_3(\Delta y_{12},\Delta y_{13},\Delta \phi_{12},\Delta \phi_{13})$
as a function of the azimuthal differences $\Delta \phi_{12}$ and 
 $\Delta \phi_{13}$,
 having fixed 
$\Delta y_{12}=\Delta y_{13}=0$. 
A quite asymmetric two-dimensional plot can be seen, resulting from the 
existence of
two correlation scales: a short-range correlation length (set by 
 single-cluster decay) 
 and a long-range correlation length (set by cluster 
formation).\footnote{Let 
us remark that this plot agrees quite well with 
the results presented in \ct{Dusling:2009ar} based on the framework of the 
glasma interpretation.}
 One can see that,
 in fact, 
 long-range azimuthal correlations mainly come from the
 $h^{\rm (3)}$ term, i.e. they are originated by   
correlated-cluster emission. 
This important point is in agreement with our main conclusion on the  
near-side ridge effect in hadronic collisions
obtained in our study of two-particle correlations 
\cite{Sanchis-Lozano:2016qda}. 
 In a 
way similar 
to two-particle azimuthal correlations 
 \cite{Sanchis-Lozano:2016qda}, 
 we conclude here that, 
in the absence of correlated-cluster emission, no long-range azimuthal 
three-particle correlations would be seen.

\begin{figure}[t!]
\begin{center}
\includegraphics[scale=0.69]{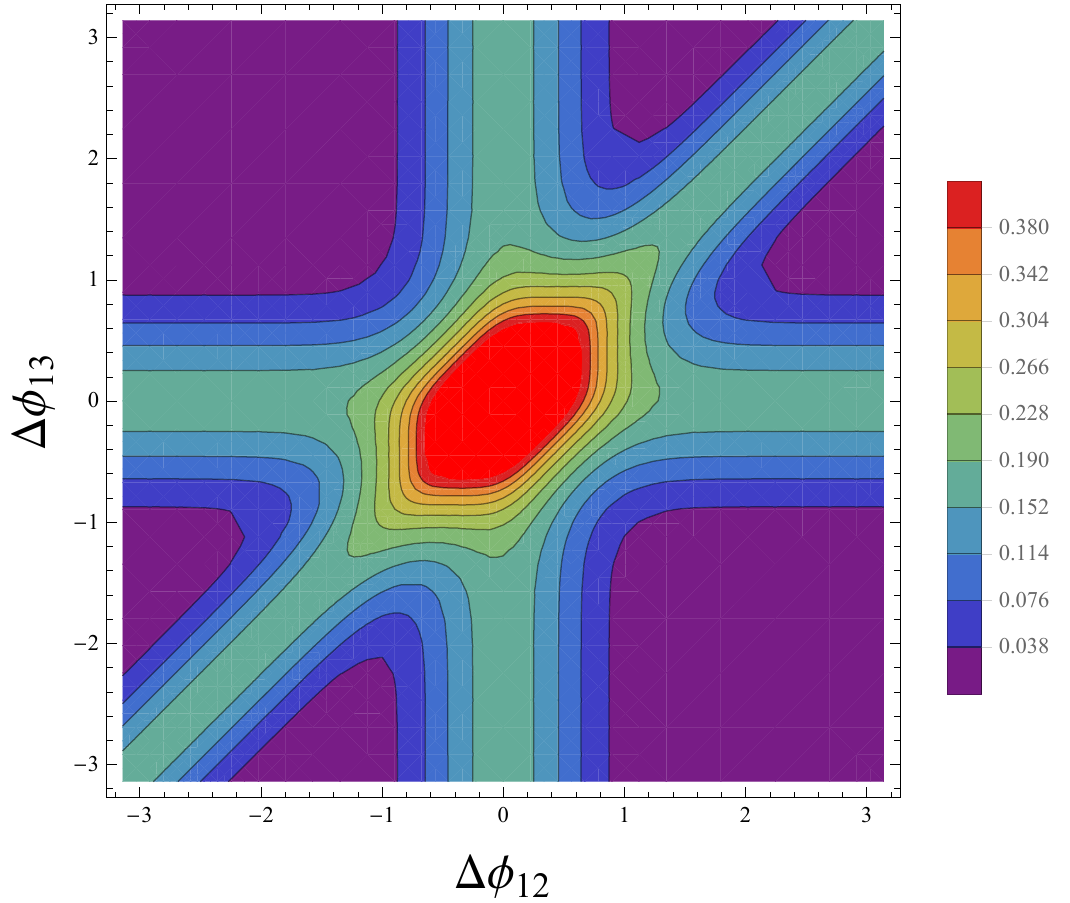}
\hspace{0.5cm}
\includegraphics[scale=0.69]{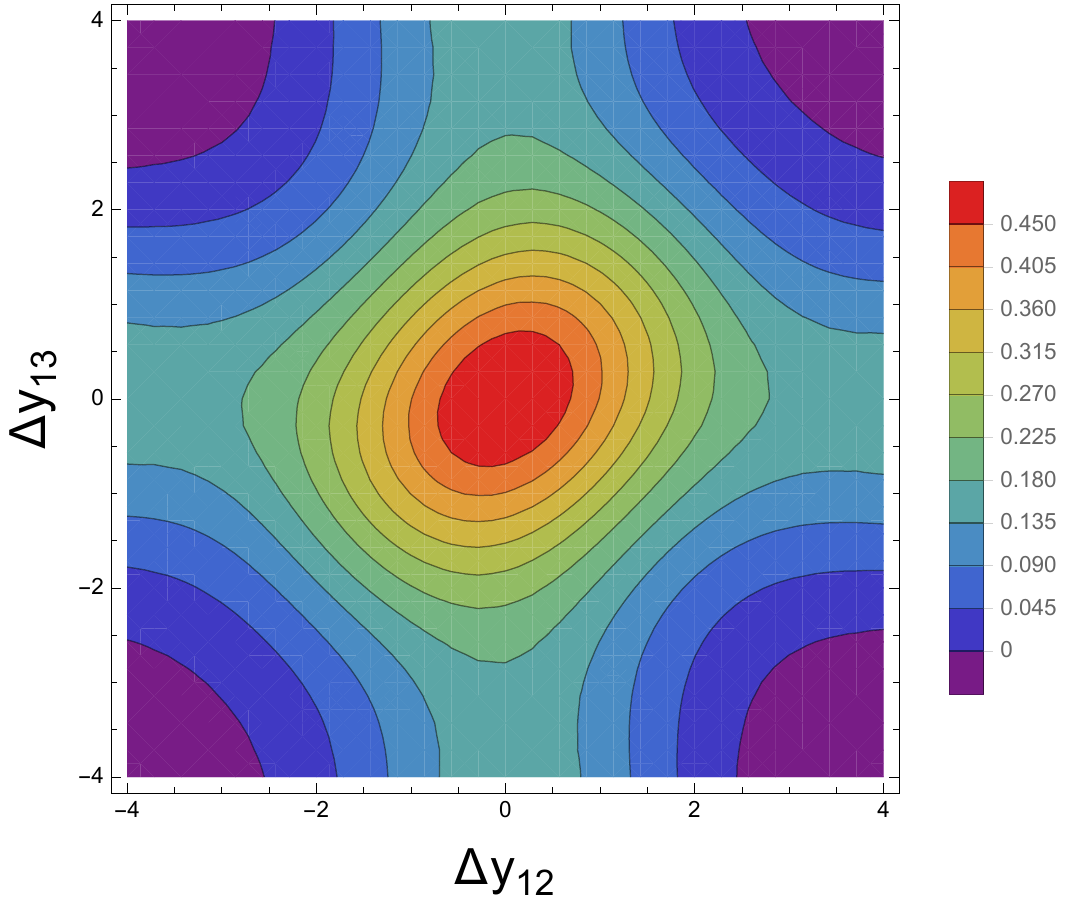}
\caption{Contour plots of $c_3(\Delta \phi_{12},\Delta 
\phi_{13})$, $\Delta y_{12}=\Delta y_{13}=0$ (left panel) 
and $c_3(\Delta 
y_{12},\Delta y_{13})$, $\Delta \phi_{12}=\Delta
\phi_{13}=0$ (right panel),  calculated using Eqs. 
 (\ref{eq:CPois})--(\ref{eq:h3}) 
 with $\delta_y=0.9$, 
$\delta_{\cl y}=4$, $\delta_{\phi}=0.14$,
$\delta_{\cl \phi}=0.5$, assuming a Poisson distribution for clusters.} 
\label{fig:contour}
\end{center}
\end{figure}

\begin{figure}[ht!]
\begin{center}
\includegraphics[scale=0.44]{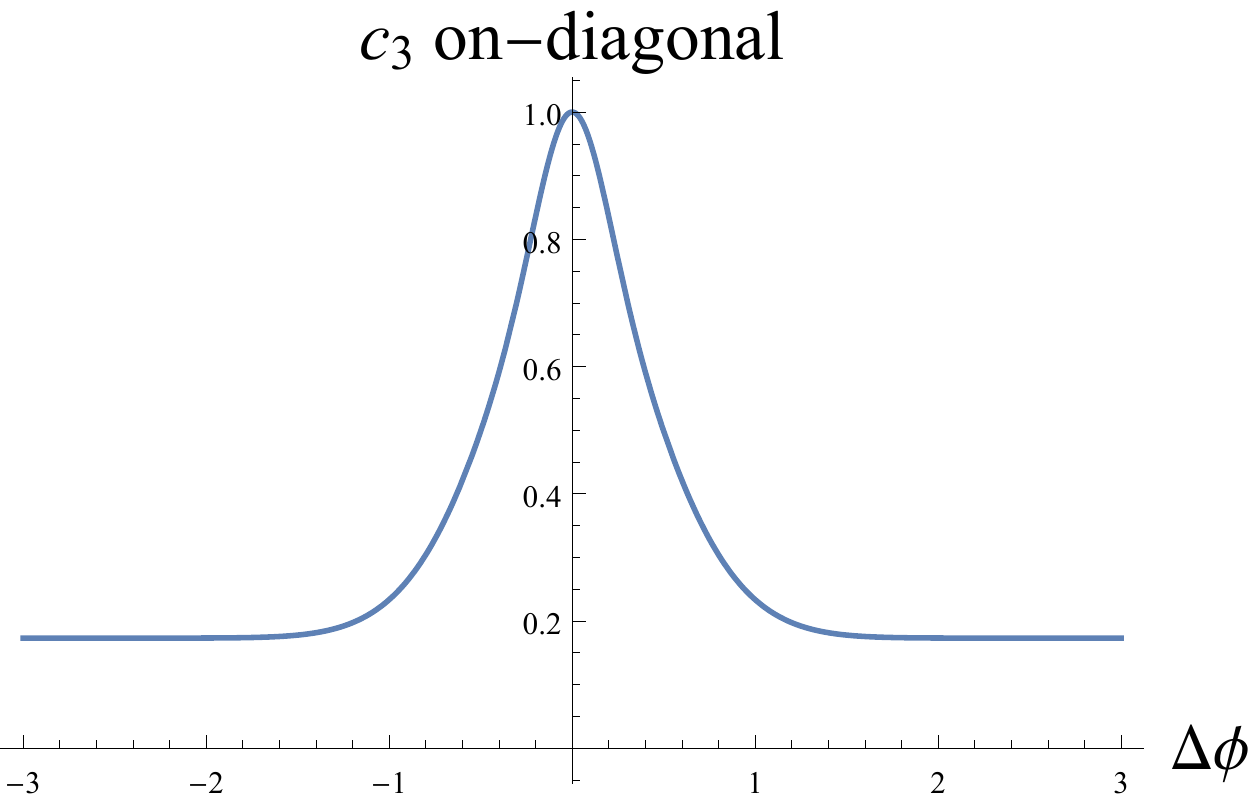}
\hspace{0.4cm}
\includegraphics[scale=0.44]{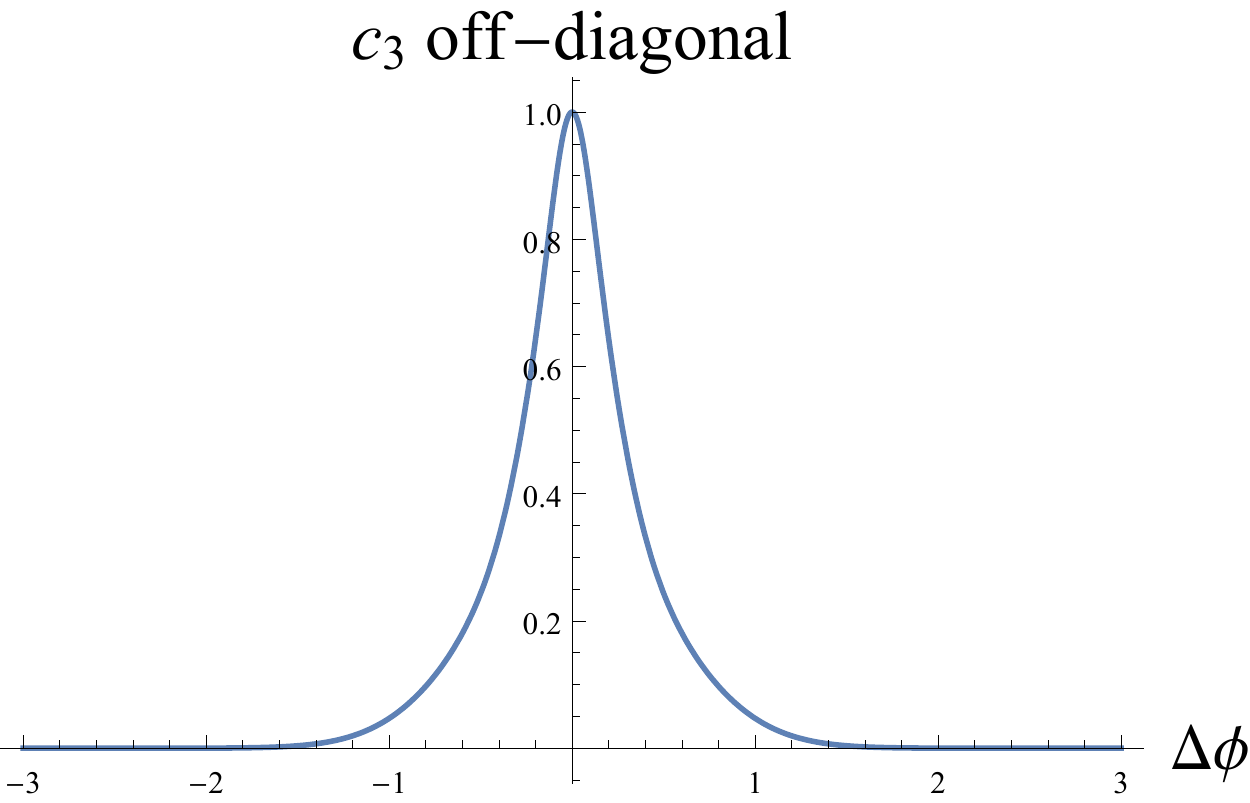}
\caption{The diagonal (left panel) and off-diagonal (right 
panel) projections of the azimuthal contour plot 
of 
$c_3(\Delta \phi_{12},\Delta \phi_{13})$ with $\Delta y_{12}=\Delta 
y_{13}=0$, shown in Fig. \ref{fig:contour}, left panel.}
\label{fig:projphi}
\end{center}
\end{figure}

\begin{figure}[ht!]
\begin{center}
\includegraphics[scale=0.44]{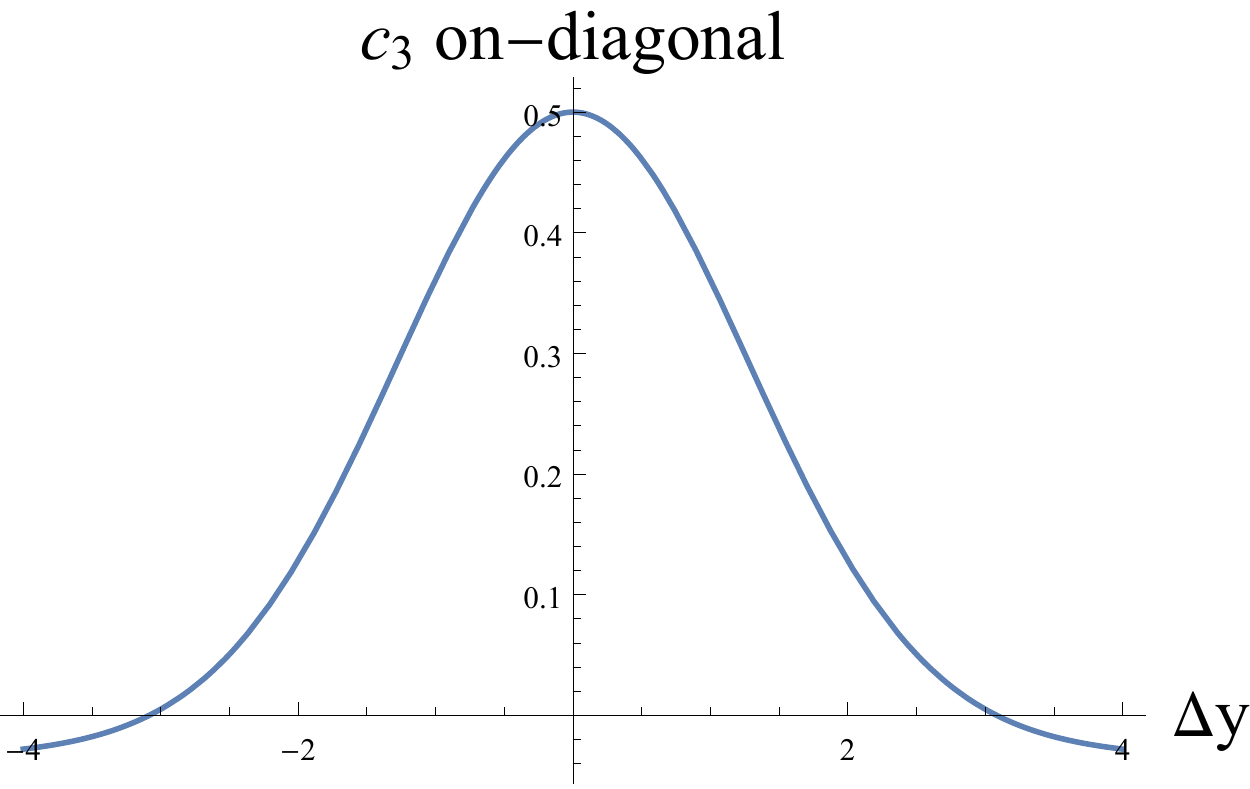}
\hspace{0.4cm}
\includegraphics[scale=0.44]{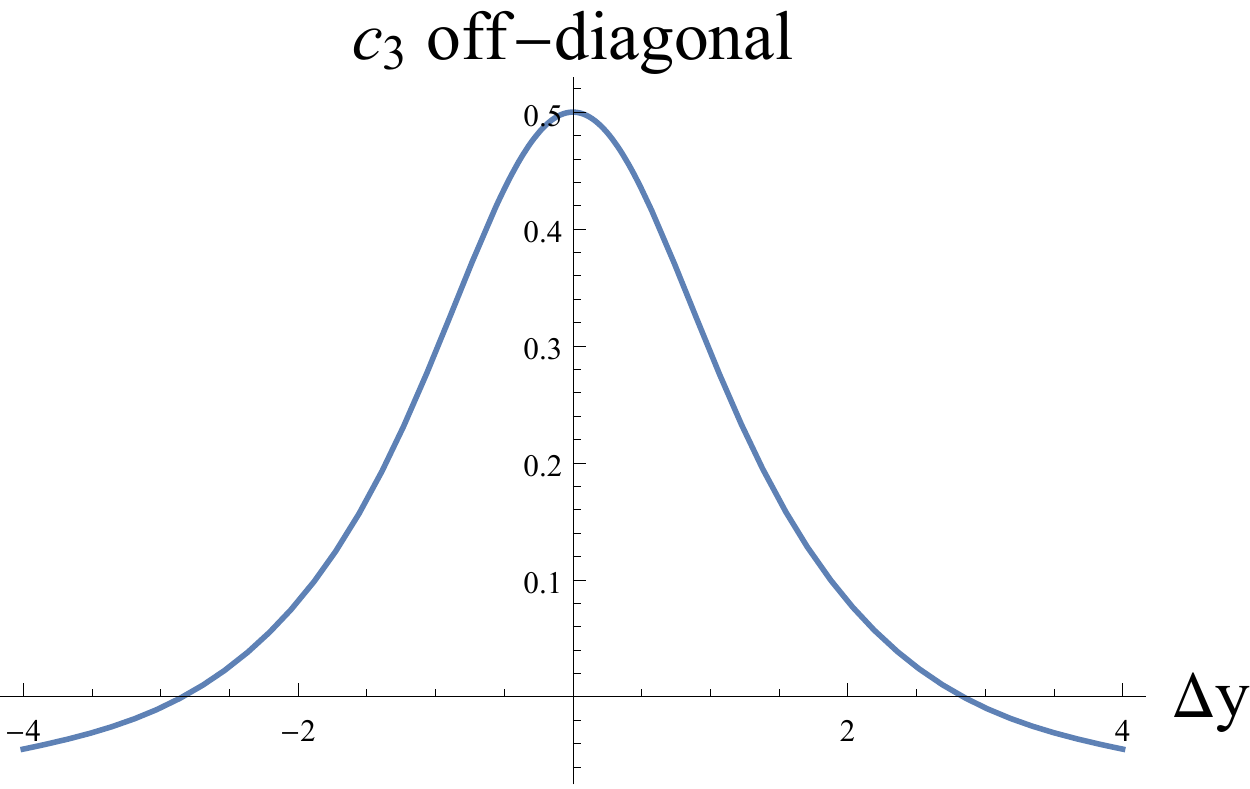}
\caption{The diagonal (left panel) and off-diagonal (right 
panel) projections of the rapidity contour plot of 
$c_3(\Delta y_{12},\Delta y_{13})$ with
 $\Delta \phi_{12}=\Delta \phi_{13}=0$,  shown in Fig. 
\ref{fig:contour}, left panel.}
\label{fig:projrap}
\end{center}
\end{figure}

 In the right panel of
  Fig. \ref{fig:contour}
 we show the contour-plot of
$c_3(\Delta y_{12},\Delta y_{13},\Delta \phi_{12},\Delta \phi_{13})$
now as a function of the rapidity differences $\Delta y_{12}$ and $\Delta y_{13}$, having fixed 
$\Delta \phi_{12}=\Delta \phi_{13}=0$. A quite different behaviour 
can be appreciated as compared to the 
azimuthal dependence on the left. Aside the central peak corresponding to
dominating short-range correlations from single clusters decays, it is 
now a rather structureless plot, in agreement with some early 
experimental measurements \cite{Netrakanti:2008jw}.

 In Fig. \ref{fig:projphi}, the projection plots of the three-particle 
correlation 
 function 
 $c_3(\Delta y_{12},\Delta y_{13},\Delta \phi_{12},\Delta \phi_{13})$
for the azimuthal-difference dependence along the diagonal 
($\Delta \phi_{12}=\Delta \phi_{13}$, left panel) and 
 off the diagonal ($\Delta \phi_{12}=-\Delta \phi_{13}$, right panel)
 are shown under the $\Delta y_{12}=\Delta y_{13}=0$ condition. 
 Again a different behaviour can be readly observed in both plots, as the 
on-diagonal correlation length
is appreciably longer than the off-diagonal correlation length. We interpret this difference 
as an indication that the former is dominated by
cluster correlations whose correlation length is larger than for particles emitted from the same
cluster that mainly populate the off-diagonal line.

 In Fig. \ref{fig:projrap}, 
the projections of the three-particle correlation 
 function 
 $c_3(\Delta y_{12},\Delta y_{13},\Delta \phi_{12},\Delta \phi_{13})$
 for the rapidity dependence only (having fixed $\Delta \phi_{12}=\Delta 
\phi_{13}=0$) 
along the diagonal ($\Delta y_{12}=\Delta y_{13}$, left plot) and 
off the diagonal ($\Delta y_{12}=-\Delta y_{13}$, right plot) are shown. 
Now, as 
expected, the two 
Gaussian-like plots 
are quite similar, reflecting that both are determined by short-distance correlations
 from single-cluster decays. 

\section{Summary} 

A study of the ridge phenomenon is presented for three-particle 
correlations, extending our previous work in the context of the 
correlated-cluster model (CCM). Gaussians are employed for azimuth 
and (pseudo)rapidity distributions, encoding 
short- and long-range correlations for clusters and final-state hadrons.
 The CCM provides a common framework to explain the ridge effect in 
proton-proton, proton-nucleus and heavy-ion collisions.  
 As obtained for two-particle correlations in our earlier study,   
 we conclude again that azimuthal correlations among clusters are 
 definitely needed
 to explain the ridge phenomenon.


\subsubsection*{Acknowledgements}

This work has been partially supported by MINECO under grant 
FPA2014-54459-P, and Generalitat Valenciana under grant 
PROMETEOII/2014/049.
 One of us (M.A.S.L.) acknowledges support from IFIC under grant 
SEV-2014-0398 of the ``Centro de Excelencia Severo Ochoa'' Programme.

\appendix

\section{Two-particle correlations}
\label{sec:corr2}

\subsection{(Pseudo)rapidity dependence}
\label{sec:corr2rap}

We will assume throughout that both clusters and particles stemming from 
clusters obey 
Gaussian distributions in rapidity space (for more detais on this Appendix 
we refer the reader to
\ct{Sanchis-Lozano:2016qda}):
\beq\label{eq:cG}
\rho^{(\cl)}(y_\cl,\phi_\cl)\ \sim\ 
\exp{\left[-\frac{y_\cl^2}{2\delta_{\cl y}^2}\right]},\ \ \ 
\rho^{(1)}(y,\phi;y_\cl,\phi_\cl)\ \sim\ 
\exp{\left[-\frac{(y-y_\cl)^2}{2\delta_y^2}\right]} \, .
\eeq

Upon integration over the cluster rapidity $y_\cl$, the $E_1^L(y)$ 
function, introduced
in Eq.(\ref{eq:rho}), reads
\beq\label{eq:E1}
E_1^L(y) \sim \int dy_\cl\ 
\exp{\left[-\frac{y_\cl^2}{2\delta_{\cl y}^2}\right]}\ 
\exp{\left[-\frac{(y-y_\cl)^2}{2\delta_y^2}\right]}\ \sim\ 
\exp{\left[-\frac{y^2}{2(\delta_y^2+\delta_{\cl y}^2)}\right]} \, .
\eeq

Hence, for two particles emitted from the two clusters one 
gets 
for 
the longitudinal
part of the 
$E_\bg$ function, introduced in Eq.(\ref{eq:rho2b}),
\beq\label{eq:E2}
E_\bg^{(2)}(y_1,y_2)\ =\ E_1^{(1)}(y_1) \cdot E_1^{(1)}(y_2)\ \sim\ 
\exp{\left[-\frac{(y_1^2+y_2^2)}{2(\delta_y^2+\delta_{\cl 
y}^2)}\right]}\, .
\eeq
Upon integration on both rapidities keeping the rapidity interval $\Delta 
y = y_1-y_2$ fixed, one gets
\beq\label{eq:2e}
e_\bg^{(2)}(\Delta y)\ \sim\ \exp{\left[-\frac{(\Delta 
y)^2}{4(\delta_y^2+\delta_{\cl y}^2)}\right]}\, ,
\eeq

For two particles stemming from the same cluster with rapidity $y_\cl$  
\[
E_\sg^{(1)}(y_1,y_2) \sim \int dy_\cl\ 
\exp{\left[-\frac{y_\cl^2}{2\delta_{\cl y}^2}\right]}
\exp{\left[-\frac{(y_1-y_\cl)^2}{2\delta_y^2}\right]}
\ \exp{\left[-\frac{(y_2-y_\cl)^2}{2\delta_y^2}\right]}
\]
\beq
\sim\ 
\exp{\left[-\frac{\delta_{\cl 
y}^2(y_1-y_2)^2}{2\delta_y^2(\delta_y^2+2\delta_{\cl y}^2)}\right]}\ 
\exp{\left[-\frac{(y_1^2+y_2^2)}{2(\delta_y^2+2\delta_{\cl y}^2)}\right]}
\, .
\eeq

After integration using the Dirac's $\delta$-function, $\delta(\Delta 
y - y_1 +y_2)$, the above 
expression leads to  
\beq
e_\sg^{(1)}(y_1,y_2)\ \sim\ 
\exp{\left[-\frac{(\Delta y)^2}{4\delta_y^2}\right]} \, .
\eeq
Notice that $\delta_{\cl y}$ drops off in the last expression so that it 
can be considered as a 
 short-range correlation contribution.

For two particles with rapidity $y_1$ and $y_2$
coming from two  (correlated) clusters with rapidities 
$y_{\cl 1}$ and $y_{\cl 2}$, respectively, we have (see 
Eqs. (\ref{eq:clustercorr}))
\[
E_\sg^{(2)}(y_1,y_2) \sim \int dy_{\cl 1}dy_{\cl 2} 
\exp{\left[-\frac{(y_{\cl 1}+y_{\cl 2})^2}{2\delta_{\cl y}^2}\right]} 
\exp{\left[-\frac{(y_1-y_{\cl 1})^2}{2\delta_y^2}\right]} 
\exp{\left[-\frac{(y_2-y_{\cl 2})^2}{2\delta_y^2}\right]}
\]
\beq
\sim\ 
\exp{\left[-\frac{(y_1+y_2)^2}{2(2\delta_y^2+\delta_{\cl y}^2)}\right]}\ .
\eeq
Using again the Dirac $\delta$-function, one gets
$e_\sg^{(2)}(\Delta y)\ \sim\ {\rm const.}$, 
which corresponds to a long-range correlation contribution. 
 
In sum, we get two pieces with different behaviours (short-range and 
long-range correlations) 
in rapidity space:
\beq\label{eq:esL}
e_\sg^{(1)}(\Delta y)\ \sim\ \exp{\left[-\frac{(\Delta 
y)^2}{4\delta_y^2}\right]}\ \ \ 
;\ \ \ e_\sg^{(2)}(\Delta y)\ \sim\  {\rm const.}
\eeq

\subsection{Azimuthal dependence}
\label{sec:corr2phi}

In addition to the hypothesis of isotropically decaying clusters in their own rest frame,
we will assume axial symmetry for cluster production in the transverse plane, i.e.
\beq
\label{eq:constphi2}
E_{\bg}^T(\phi_1,\phi_2)\ \sim\ \mathrm{const.}\ \to\ e_{\bg}^T(\Delta 
\phi)\ \sim\ \mathrm{const.}\
\eeq

Thus, the distribution for two particles, emitted from the same cluster 
with azimuthal angle $\phi_\cl$ should obey 
\beq
\int d\phi_\cl\ 
\exp{\left[-\frac{(\phi_1-\phi_\cl)^2}{2\delta_{\phi}^2}\right]}\ 
\exp{\left[-\frac{(\phi_2-\phi_\cl)^2}{2\delta_{\phi}^2}\right]}\ 
\sim\ \exp{\left[-\frac{(\phi_1-\phi_2)^2}{4\delta_{\phi}^2}\right]}
\eeq
for small azimuthal angles. Therefore, regarding the azimuthal dependence we can write
\beq
e_\sg^{(1)}(\Delta \phi)\ \sim\ \exp{\left[-\frac{(\Delta 
\phi)^2}{4\delta_{\phi}^2}\right]} \, .
\eeq
 As for the above rapidity correlations, the dependence on 
$\delta_{\cl \phi}$ drops off in this expression, so that it can be 
referred as a short-range correlation term. 

On the other hand, we will assume that 
clusters are produced in a correlated way according
to Eqs.(\ref{eq:clustercorr}). Hence 
for two particles with azimuthal angles $\phi_1$ and $\phi_2$
coming from {two (correlated) clusters with azimuthal angles 
$\phi_{\cl 
1}$ 
and $\phi_{\cl 2}$, 
 we will write
\beq
E_\sg^T(\phi_1,\phi_2) \sim \int d\phi_{\cl 1}d\phi_{\cl 2} 
\exp{\left[-\frac{(\phi_{\cl 
1}-\phi_{\cl 2})^2}{2\delta_{\cl \phi}^2}\right]} 
\exp{\left[-\frac{(\phi_1-\phi_{\cl 1})^2}{2\delta_{\phi}^2}\right]} 
\exp{\left[-\frac{(\phi_2-\phi_{\cl 2})^2}{2\delta_{\phi}^2}\right]}
\eeq
\[
\sim\ 
\exp{\left[-\frac{(\phi_1-\phi_2)^2}{2(2\delta_{\phi}^2+
\delta_{\cl \phi}^2)}\right]} \, ,
 \] 
that directly leads to
\beq
e_\sg^{(2)}(\Delta \phi)\ 
\sim\ \exp{\left[-\frac{(\Delta \phi)^2}{2(2\delta_{\phi}^2+\delta_{\cl 
\phi}^2)}\right]} \, ,
\eeq
 which corresponds to a long-range correlation contribution.

\subsection{Final expressions}
\label{sec:final2}

 In sum, we find
  that the short-range and long-range pieces of the $e_\sg(\Delta 
y,\Delta \phi)$ function
can be written as
\[
e_\sg^{(1)}(\Delta y, \Delta \phi)\ \sim\ \exp{\left[-\frac{(\Delta 
y)^2}{4\delta_y^2}\right]}\ 
\exp{\left[-\frac{(\Delta \phi)^2}{4\delta_{\phi}^2}\right]}
\]
and
\[
e_\sg^{(2)}(\Delta y, \Delta \phi)\  \sim\ 
\exp{\left[-\frac{(\Delta \phi)^2}{2(2\delta_{\phi}^2+\delta_{\cl 
\phi}^2)}\right]} \, .
\]

 Note that
$e_\bg(\Delta y, \Delta \phi)$ only retains dependence
on the rapidity variable for isotropic cluster production in the transverse 
plane,
 \[
e_\bg(\Delta y, \Delta \phi)\ \sim\ 
\exp{\left[-\frac{(\Delta y)^2}{4(\delta_y^2+\delta_{\cl y}^2)}\right]} \, 
.
\]

\section{Three-particle correlations}
\label{sec:corr3}

\subsection{(Pseudo)rapidity dependence}
\label{sec:cor2phi}



Similarly to the above two-particle calculations in Appendix 
\ref{sec:corr2rap}, for 
three particles emitted from three clusters one gets 
for 
the longitudinal
part of the 
$E_\bg$ function,
\beq\label{eq:E3}
E_\bg^{(3)}(y)\ =\ E_1^L(y_1) \cdot E_1^L(y_2) \cdot E_1^L(y_3)\sim\ 
\exp{\left[-\frac{(y_1^2+y_2^2+y_3^2)}{2(\delta_y^2+\delta_{\cl 
y}^2)}\right]}\, ,
\eeq
where 
$E_1^L(y)$ is defined in Eq.
(\ref{eq:E1}).
Upon integration over all three rapidities and keeping the rapidity 
intervals 
$\Delta y_{12} = y_1-y_2$ and $\Delta y_{13} = y_1-y_3$ fixed, one gets
\beq\label{eq:3e}
e_\bg^{(3)}(\vec{\Delta y})\ \sim\ 
\exp{\left[-\frac{(\Delta y_{12})^2+(\Delta y_{13})^2-\Delta y_{12}\Delta 
y_{13}}{3(\delta_y^2+\delta_{\cl y}^2)}\right]}\, .
\eeq
Since $\Delta y_{23}=\Delta y_{13}-\Delta y_{12}$, the above 
equation can be also expressed in terms of all three rapidity intervals
\beq\label{eq:3ebis}
e_\bg^{(3)}(\vec{\Delta y})\ \sim\ \exp{\left[-
\frac{(\Delta y_{12})^2+(\Delta y_{13})^2+(\Delta 
y_{23})^2}{6(\delta_y^2+\delta_{\cl y}^2)}\right]}\, .
\eeq

For three particles stemming from the same cluster with rapidity 
$y_\cl$,  
\[
E_\sg^{(1)}(\vec{y}) \sim \int dy_\cl\ 
\exp{\left[-\frac{y_\cl^2}{2\delta_{\cl y}^2}\right]}\ 
\exp{\left[-\frac{(y_1-y_\cl)^2+(y_2-y_\cl)^2+(y_3-y_\cl)^2}{2\delta_y^2}\right]}
\]
\beq
\sim\ 
\exp{\left[-\frac{\delta_{\cl 
y}^2(y_1^2+y_2^2+y_3^2-y_1y_2-y_1y_3-y_2y_3)}{2\delta_y^2(\delta_y^2+3\delta_{\cl 
y}^2)}\right]}\ 
\exp{\left[-\frac{(y_1^2+y_2^2+y_3^2)}{2(\delta_y^2+3\delta_{\cl 
y}^2)}\right]}
\, .
\eeq

After integration using the Dirac's $\delta$-functions (see 
Eq.(\ref{eq:dirac3}), the above 
expression leads to  
\beq
e_\sg^{(1)}(\vec{\Delta y})\ \sim\ 
\exp{\left[-\frac{(\Delta y_{12})^2+(\Delta y_{13})^2-\Delta y_{12}\Delta 
y_{13}}{3\delta_y^2}\right]} \, .
\eeq
Notice again that $\delta_{\cl y}$ drops off in the last expression so 
that it 
can be referred to as a 
short-range correlation contribution.

Again, using the $\vec{\Delta y}$ interdependence, the above equation can 
be 
expressed as 
\beq
e_\sg^{(1)}(\vec{\Delta y})\ \sim\ 
\exp{\left[-\frac{(\Delta y_{12})^2+(\Delta y_{13})^2+(\Delta 
y_{23})^2}{6\delta_y^2}\right]}\, .
\eeq

For three particles coming from two clusters emitted in a 
correlated way (see Eqs. (\ref{eq:clustercorr})), we have three 
 possibilities:
\[
E_\sg^{(2)}(\vec{y}) \sim \int dy_{\cl 1}dy_{\cl 2} 
\exp{\left[-\frac{(y_{\cl 1}+y_{\cl 2})^2}{2\delta_{\cl y}^2}\right]}
\exp{\left[-\frac{(y_1-y_{\cl 1})^2+(y_2-y_{\cl 2})^2+(y_3-y_{\cl 
2})^2}{2\delta_y^2}\right]}
\]
\beq
\sim\ 
\exp{\left[-\frac{\delta_{\cl 
y}^2(y_2-y_3)^2+2\delta_y^2(y_1^2+y_2^2+y_3^2+y_1y_2+y_1y_3-y_2y_3)}
{2\delta_y^2(3\delta_y^2+2\delta_{\cl y}^2)}\right]}\, ,
\eeq
\[
E_\sg^{(2)}(\vec{y}) \sim \int dy_{\cl 1}dy_{\cl 2} 
\exp{\left[-\frac{(y_{\cl 1}+y_{\cl 2})^2}{2\delta_{\cl y}^2}\right]}
\exp{\left[-\frac{(y_2-y_{\cl 1})^2+(y_1-y_{\cl 2})^2+(y_3-y_{\cl 
2})^2}{2\delta_y^2}\right]}\, ,
\]
\beq
\sim\ 
\exp{\left[-\frac{\delta_{\cl 
y}^2(y_1-y_3)^2+2\delta_y^2(y_1^2+y_2^2+y_3^2+y_1y_2+y_2y_3-y_1y_3)}
{2\delta_y^2(3\delta_y^2+2\delta_{\cl y}^2)}\right]}\, ,
\eeq
\[
E_\sg^{(2)}(\vec{y}) \sim \int dy_{\cl 1}dy_{\cl 2} 
\exp{\left[-\frac{(y_{\cl 1}+y_{\cl 2})^2}{2\delta_{\cl y}^2}\right]}
\exp{\left[-\frac{(y_3-y_{\cl 1})^2+(y_1-y_{\cl 2})^2+(y_2-y_{\cl 
2})^2}{2\delta_y^2}\right]}
\]
\beq
\sim\ 
\exp{\left[-\frac{\delta_{\cl 
y}^2(y_1-y_2)^2+2\delta_y^2(y_1^2+y_2^2+y_3^2+y_1y_3+y_2y_3-y_1y_2)}
{2\delta_y^2(3\delta_y^2+2\delta_{\cl y}^2)}\right]}\ .
\eeq

Using again the Dirac's $\delta$-functions (\ref{eq:dirac3}), one gets 
respectively
\beq
e_\sg^{(2)}(\vec{\Delta y})\ \sim\ 
\exp{\left[-\frac{(\Delta y_{23})^2}{4\delta_y^2}\right]}\ =\ 
\exp{\left[-\frac{(\Delta y_{12})^2+(\Delta y_{13})^2-2\Delta y_{12}\Delta 
y_{13}}{4\delta_y^2}\right]}\, ,
\eeq

\beq
e_\sg^{(2)}(\vec{\Delta y})\ \sim\ 
\exp{\left[-\frac{(\Delta y_{13})^2}{4\delta_y^2}\right]}\, ,
\eeq
and
\beq
e_\sg^{(2)}(\vec{\Delta y})\ \sim\ 
\exp{\left[-\frac{(\Delta y_{12})^2}{4\delta_y^2}\right]}\, .
\eeq

Similar to a single-cluster case, the $\delta_{\cl y}$ drops off in the 
these expressions which therefore 
can be referred to as other 
short-range correlation contributions.

For three particles with rapidities $y_1$, $y_2$ and $y_3$
coming from three (correlated) clusters with rapidities 
$y_{\cl 1}$, $y_{\cl 2}$ and $y_{\cl 3}$, respectively, we have (see Eqs. 
(\ref{eq:clustercorr})):
\[
E_\sg^{(3)}(\vec{y}) \sim \int dy_{\cl 1 }dy_{\cl 2}dy_{\cl 3}
\exp{\left[-\frac{(y_{\cl 1}+y_{\cl 2}+y_{\cl 3})^2}{2\delta_{\cl 
y}^2}\right]} 
\exp{\left[-\frac{(y_1-y_{\cl 1})^2+(y_2-y_{\cl 2})^2+(y_3-y_{\cl 
3})^2}{2\delta_y^2}\right]}
\]
\beq
\sim\ \exp{\left[-\frac{(y_1+y_2+y_3)^2}{2(3\delta_y^2+\delta_{\cl 
y}^2)}\right]}\ .
\eeq

As commented in the main text, the Gaussian depending on 
the sum of rapidities $y_{\cl 1}+y_{\cl 2}+y_{\cl 3}$
stems from the requirement of partial (longitudinal) momentum conservation. It
takes into account different topologies for cluster emission once
integrated upon their rapidities. 

Applying again the Dirac's $\delta$-functions, one gets
\beq
e_\sg^{(3)}(\vec{\Delta y})\ \sim\ {\rm const.}\, ,
\eeq 
which corresponds to a long-range correlations contribution in rapidity 
phase-space.

 
In sum, we get several pieces with different behaviours  
in rapidity space:
\beq\label{eq:esL3}
e_\sg^{(1)}(\vec{\Delta y})\ \sim\ 
\exp{\left[-\frac{(\Delta y_{12})^2+(\Delta y_{13})^2-\Delta y_{12}\Delta 
y_{13}}{3\delta_y^2}\right]}\, ,
\eeq
\beq
e_\sg^{(2)}(\vec{\Delta y}) \sim 
 \exp{\left[-\frac{(\Delta y_{12})^2+(\Delta y_{13})^2-2 \Delta 
y_{12}\Delta y_{13}}{4\delta_y^2}\right]}+
\exp{\left[-\frac{(\Delta 
y_{12})^2}{4\delta_y^2}\right]}+\exp{\left[-\frac{(\Delta 
y_{13})^2}{4\delta_y^2}\right]}\, ,
\eeq
\beq
\ \ e_\sg^{(3)}(\vec{\Delta y})\ \sim\  {\rm const.}
\eeq

\subsection{Azimuthal dependence}
\label{sec:corr3phi}


The distribution for three particles, emitted from the same cluster 
with azimuthal angle $\phi_\cl$ should obey 
\beq
\int d\phi_\cl\ 
\exp{\left[-\frac{(\phi_1-\phi_\cl)^2}{2\delta_{\phi}^2}\right]}\ 
\exp{\left[-\frac{(\phi_2-\phi_\cl)^2}{2\delta_{\phi}^2}\right]}\ 
\exp{\left[-\frac{(\phi_3-\phi_\cl)^2}{2\delta_{\phi}^2}\right]} 
\eeq
for small azimuthal differences. Therefore, regarding the azimuthal dependence we can write
\beq
e_\sg^{(1)}(\vec{\Delta \phi})\ \sim\ 
\exp{\left[-\frac{(\Delta \phi_{12})^2+(\Delta \phi_{13})^2-\Delta 
\phi_{12}\Delta \phi_{13}}
{3\delta_{\phi}^2}\right]}=
\exp{\left[-\frac{(\Delta \phi_{12})^2+ (\Delta \phi_{13})^2+(\Delta 
\phi_{23})^2}
{6\delta_{\phi}^2}\right]}\, .
\eeq

For three particles with azimuthal angles $\phi_1$ and $\phi_2$
coming from two  clusters 
with azimuthal angles $\phi_{\cl 1}$ 
and $\phi_{\cl 2}$ emitted in a correlated way (see Eqs. 
(\ref{eq:clustercorr})), three possible topologies are:
\beq
E_\sg^{(2)}(\vec{\phi}) \sim \int d\phi_{\cl 1}d\phi_{\cl 2} 
\exp{\left[-\frac{(\phi_{\cl 1}-\phi_{\cl 2})^2}{2\delta_{\cl 
\phi}^2}\right]}\ 
\exp{\left[-\frac{(\phi_1-\phi_{\cl 1})^2+(\phi_2-\phi_{\cl 
2})^2+(\phi_3-\phi_{\cl 2})^2}{2\delta_{\phi}^2}\right]} \, ,
 \eeq
and
\beq
E_\sg^{(2)}(\vec{\phi}) \sim \int d\phi_{\cl 1}d\phi_{\cl 2} 
\exp{\left[-\frac{(\phi_{\cl 1}-\phi_{\cl 2})^2}{2\delta_{\cl 
\phi}^2}\right]}\
\exp{\left[-\frac{(\phi_2-\phi_{\cl 1})^2+(\phi_1-\phi_{\cl 
2})^2+(\phi_3-\phi_{\cl 2})^2}{2\delta_{\phi}^2}\right]} \, ,
 \eeq
and
\beq
E_\sg^{(2)}(\vec{\phi}) \sim \int d\phi_{\cl 1}d\phi_{\cl 2} 
\exp{\left[-\frac{(\phi_{\cl 1}-\phi_{\cl 2})^2}{2\delta_{\cl 
\phi}^2}\right]}\ 
\exp{\left[-\frac{(\phi_3-\phi_{\cl 1})^2+(\phi_1-\phi_{\cl 
2})^2+(\phi_2-\phi_{\cl 2})^2}{2\delta_{\phi}^2}\right]} \, .
 \eeq
 
The above integrals lead to
\[
E_\sg^{(2)}(\vec{\phi})\ 
\sim\ \exp{\left[-\frac{\delta_{\cl \phi}^2(\phi_2-\phi_3)^2
+2\delta_{\phi}^2(\phi_1^2+\phi_2^2+
\phi_3^2-\phi_1\phi_2-\phi_1\phi_3-\phi_2\phi_3)}
{2\delta_{\phi}^2(3\delta_{\phi}^2+2\delta_{\cl \phi}^2)}\right]} \, ,
\]
\[
E_\sg^{(2)}(\vec{\phi})\ 
\sim\ \exp{\left[-\frac{\delta_{\cl \phi}^2(\phi_1-\phi_3)^2
+2\delta_{\phi}^2(\phi_1^2+\phi_2^2+\phi_3^2-\phi_1\phi_2-\phi_1\phi_3-\phi_2\phi_3)}
{2\delta_{\phi}^2(3\delta_{\phi}^2+2\delta_{\cl \phi}^2)}\right]} \, ,
\]
\[
E_\sg^{(2)}(\vec{\phi})\ 
\sim\ \exp{\left[-\frac{\delta_{\cl \phi}^2(\phi_1-\phi_3)^2
+2\delta_{\phi}^2(\phi_1^2+\phi_2^2+(\phi_3)^2-\phi_1\phi_2-\phi_1\phi_3-\phi_2\phi_3)}
{2\delta_{\phi}^2(3\delta_{\phi}^2+2\delta_{\cl \phi}^2)}\right]} \, ,
\]
which can be rewritten as
\beq
e_\sg^{(2)}(\vec{\Delta \phi})\ 
\sim\ \exp{\left[-\frac{\delta_{\cl \phi}^2(\Delta \phi_{23})^2}
{2\delta_{\phi}^2(3\delta_{\phi}^2+2\delta_{\cl \phi}^2)}\right]}\
\exp{\left[-\frac{(\Delta \phi_{12})^2+ (\Delta \phi_{13})^2+(\Delta 
\phi_{23})^2}
{2(3\delta_{\phi}^2+2\delta_{\cl \phi}^2)}\right]} \, ,
\eeq
\beq
e_\sg^{(2)}(\vec{\Delta \phi})\ 
\sim\ \exp{\left[-\frac{\delta_{\cl \phi}^2(\Delta \phi_{13})^2}
{2\delta_{\phi}^2(3\delta_{\phi}^2+2\delta_{\cl \phi}^2)}\right]}\
\exp{\left[-\frac{(\Delta \phi_{12})^2+ (\Delta \phi_{13})^2+(\Delta 
\phi_{23})^2}
{2(3\delta_{\phi}^2+2\delta_{\cl \phi}^2)}\right]} \, ,
\eeq
\beq
e_\sg^{(2)}(\vec{\Delta \phi})\ 
\sim\ \exp{\left[-\frac{\delta_{\cl \phi}^2(\Delta \phi_{12})^2}
{2\delta_{\phi}^2(3\delta_{\phi}^2+2\delta_{\cl \phi}^2)}\right]}\
\exp{\left[-\frac{(\Delta \phi_{12})^2+ (\Delta \phi_{13})^2+(\Delta 
\phi_{23})^2}
{2(3\delta_{\phi}^2+2\delta_{\cl \phi}^2)}\right]} \, .
\eeq

For three clusters with azimuthal angles $\phi_{\cl 1}$, 
$\phi_{\cl 2}$ and 
$\phi_{\cl 3}$, all of them emitted in a correlated way one has (see Eqs. 
(\ref{eq:clustercorr})):
\[
E_\sg^{(3)}(\vec{\phi}) \sim \int d\vec{\phi}_{\cl} 
\exp{\left[-\frac{(\phi_{\cl 1}-\phi_{\cl 2})^2+(\phi_{\cl 1}-\phi_{\cl 
3})^2+(\phi_{\cl 2}-\phi_{\cl 3})^2}{2\delta_{\cl \phi}^2}\right]}
\]
\[
\times\
\exp{\left[-\frac{(\phi_1-\phi_{\cl 1})^2+(\phi_2-\phi_{\cl 
2})^2+(\phi_3-\phi_{\cl 3})^2}{2\delta_{\phi}^2}\right]}
\]
\beq
\sim\ 
\exp{\left[-\frac{\phi_1^2+\phi_2^2+\phi_3^2-\phi_1\phi_2-\phi_1\phi_3-\phi_2\phi_3}
{3\delta_{\phi}^2+\delta_{\cl \phi}^2}\right]}\, ,
\eeq
which can be rewritten as
\beq
e_\sg^{(3)}(\vec{\Delta \phi})\ \sim\  
\exp{\left[-\frac{(\Delta \phi_{12})^2+(\Delta \phi_{13})^2-\Delta 
\phi_{12} \Delta \phi_{13}}
{3\delta_{\phi}^2+\delta_{\cl \phi}^2}\right]}\ \sim\ 
\exp{\left[-\frac{(\Delta \phi_{12})^2+(\Delta \phi_{13})^2+(\Delta 
\phi_{23})^2}
{2(3\delta_{\phi}^2+\delta_{\cl \phi}^2)}\right]}\, .
\eeq

For two clusters (out of three) emitted in a correlated way, the three 
possible topologies are:
\[
E_\sg^{(3)}(\vec{\phi}) \sim \int d\vec{\phi}_{\cl} 
\exp{\left[-\frac{(\phi_{\cl 2}-\phi_{\cl 3})^2}{2\delta_{\cl 
\phi}^2}\right]} 
\exp{\left[-\frac{(\phi_1-\phi_{\cl 1})^2+(\phi_2-\phi_{\cl 
2})^2+(\phi_3-\phi_{\cl 3})^2}{2\delta_{\phi}^2}\right]}
\]
\[
\sim\
\exp{\left[-\frac{(\phi_2-\phi_3)^2}
{2(2\delta_{\phi}^2+\delta_{\cl \phi}^2)}\right]}\, , 
\]
which can be rewritten as
\beq
e_\sg^{(3)}(\vec{\Delta \phi)}\ 
\sim\ \exp{\left[-\frac{(\Delta \phi_{12})^2+(\Delta \phi_{13})^2-2\Delta 
\phi_{12}\Delta \phi_{13}}
{2(2\delta_{\phi}^2+\delta_{\cl \phi}^2)}\right]}\ \sim\ 
\exp{\left[-\frac{(\Delta \phi_{23})^2}
{2(2\delta_{\phi}^2+\delta_{\cl \phi}^2)}\right]}\, ;
\label{eq:c2of3phi}
\eeq

 \[
E_\sg^{(3)}(\vec{\phi}) \sim \int d\vec{\phi}_{\cl} 
\exp{\left[-\frac{(\phi_{\cl 1}-\phi_{\cl 2})^2}{2\delta_{\cl 
\phi}^2}\right]}
\exp{\left[-\frac{(\phi_1-\phi_{\cl 1})^2+(\phi_2-\phi_{\cl 
2})^2+(\phi_3-\phi_{\cl 3})^2}{2\delta_{\phi}^2}\right]}
\]
\[
\sim\ 
\exp{\left[-\frac{(\Delta \phi_{12})^2}
{2(2\delta_{\phi}^2+\delta_{\cl \phi}^2)}\right]}\ ,
\]
\[
E_\sg^{(3)}(\vec{\phi}) \sim \int d\vec{\phi}_{\cl} 
\exp{\left[-\frac{(\phi_{\cl 1}-\phi_{\cl 3})^2}{2\delta_{\cl 
\phi}^2}\right]}
\exp{\left[-\frac{(\phi_1-\phi_{\cl 1})^2+(\phi_2-\phi_{\cl 
2})^2+(\phi_3-\phi_{\cl 3})^2}{2\delta_{\phi}^2}\right]}
\]
\[
\sim\ 
\exp{\left[-\frac{(\Delta \phi_{13})^2}
{2(2\delta_{\phi}^2+\delta_{\cl \phi}^2)}\right]}\ .
\]

Notice that, conversely to the (pseudo)rapidity dependence, we consider
pairwise azimuthal correlations among clusters as 
potential contributions to the ridge effect.
 Therefore, no two-cluster contributions similar to the three ones just 
 above 
 to appear for pseudorapidities.

For three clusters, all of them independently emitted, one gets
\[
E_\sg^{(3)}(\vec{\phi}) \sim \int d\phi_{\cl 1}d\phi_{\cl 2}d\phi_{\cl 3}\ 
\exp{\left[-\frac{(\phi_1-\phi_{\cl 1})^2+(\phi_2-\phi_{\cl 
2})^2+(\phi_3-\phi_{\cl 2})^2}{2\delta_{\phi}^2}\right]}\ 
\sim {\rm const.} \, ,
\]
leading to
\beq
e_\sg^{(3)}(\vec{\Delta \phi})\ \sim\ {\rm const.}
\eeq

\newpage

\subsection{Final expressions}
\label{sec:final3}

Taking into account 
that in the transverse direction,
 in addition to the hypothesis of isotropically decaying clusters in 
 their own rest frame,
 we assume axial symmetry for cluster production 
 i.e.
 \beq
 E_{\bg}^T(\phi_1,\phi_2,\phi_3)\ \sim\ \mathrm{const.}\ \to\ 
 e_{\bg}^T(\vec{\Delta \phi})\ \sim\ \mathrm{const.}\, , 
 \eeq
similarly to two-particle correlations, Eq. (\ref{eq:constphi2}),
 the final experssions are as follows: 
\\

\noi
{\it - for one cluster}:

\beq\label{eq:h1bis}
h^{\rm (1)}(\vec{\Delta y},\vec{\Delta \phi})=\frac{e_s^{(1)}}{e_\bg^{(3)}} 
\eeq
\[
\sim\ 
\exp{\left[-\frac{\delta_{\cl y}^2\{(\Delta y_{12})^2+(\Delta 
y_{13})^2+(\Delta y_{23})^2\}}
{6\delta_y^2(\delta_y^2+\delta_{\cl y}^2)}
\right]}
\exp{\left[-\frac{(\Delta \phi_{12})^2+(\Delta \phi_{13})^2+(\Delta 
\phi_{23})^2}{6\delta_{\phi}^2}\right]}\, ,
\]
\\

\noi
{\it - for two clusters:}\\

\beq\label{eq:h2bis}
h^{\rm (2)}(\vec{\Delta y},\vec{\Delta \phi})=\frac{e_s^{(2)}}{e_\bg^{(3)}} 
\sim 
\left( 
\exp{\left[-\frac{(\Delta y_{12})^2}{4\delta_y^2}\right]}+
\exp{\left[-\frac{(\Delta y_{13})^2}{4\delta_y^2}\right]}+
\exp{\left[-\frac{(\Delta y_{23})^2}{4\delta_y^2}\right]}
\right)
\eeq
\[
\times\
\exp{\left[\frac{(\Delta y_{12})^2+(\Delta y_{13})^2+(\Delta y_{23})^2}
{6(\delta_y^2+\delta_{\cl y}^2)}\right]} 
\]
\[
\times\
\left(
\exp{\left[-\frac{\delta_{\cl \phi}^2(\Delta 
\phi_{12})^2}{2\delta_{\phi}^2(3\delta_{\phi}^2+2\delta_{\cl 
\phi}^2)}\right]}+
\exp{\left[-\frac{\delta_{\cl \phi}^2(\Delta 
\phi_{13})^2}{2\delta_{\phi}^2(3\delta_{\phi}^2+2\delta_{\cl \phi}^2)}\right]}+ 
\exp{\left[-\frac{\delta_{\cl \phi}^2(\Delta 
\phi_{23})^2}{2\delta_{\phi}^2(3\delta_{\phi}^2+2\delta_{\cl 
\phi}^2)}\right]}
\right) 
\]
\[
\times\
\exp{\left[-\frac{(\Delta \phi_{12})^2+(\Delta \phi_{13})^2+(\Delta 
\phi_{23})^2}
{2(3\delta_{\phi}^2+2\delta_{\cl \phi}^2)}\right]}
\, ,
\]
\\

\noi
{\it - for three clusters:}

\beq\label{eq:h3bis}
h^{\rm (3)}(\vec{\Delta y},\vec{\Delta \phi})=
\frac{e_s^{(3)}}{e_\bg^{(3)}}\sim  
\exp{\left[\frac{(\Delta y_{12})^2+(\Delta y_{13})^2+(\Delta 
y_{23})^2}{6(\delta_y^2+\delta_{\cl 
y}^2)}\right]}
\eeq
\[
\times\ 
\left(
\exp{\left[-\frac{(\Delta \phi_{12})^2+(\Delta \phi_{13})^2+(\Delta 
\phi_{23})^2}{2(3\delta_{\phi}^2+\delta_{\cl \phi}^2)}\right]}\right.
\]
\[
\left. +\ \exp{\left[-\frac{(\Delta 
\phi_{12})^2}{2(2\delta_{\phi}^2+\delta_{\cl \phi}^2)}\right]}+
\exp{\left[-\frac{(\Delta 
\phi_{13})^2}{2(2\delta_{\phi}^2+\delta_{\cl \phi}^2)}\right]}+ 
\exp{\left[-\frac{(\Delta 
\phi_{23})^2}{2(2\delta_{\phi}^2+\delta_{\cl \phi}^2)}\right]} \right)\, .
\]

\end{document}